# Non-invasive Scanning Raman Spectroscopy and Tomography for Graphene Membrane Characterization

Stefan Wagner[1], Thomas Dieing[2], Alba Centeno[3], Amaia Zurutuza[3], Anderson D. Smith[4], Mikael Östling[4], Satender Kataria[1], Max C. Lemme[1]

[1]Department of Electrical Engineering and Computer Science, University of Siegen, Hölderlinstrasse 3, 57076 Siegen, Germany
[2]WITec Wissenschaftliche Instrumente und Technologie GmbH, Lise-Meitner-Str. 6, 89081 Ulm, Germany
[3]Graphenea S.A., Avenida de Tolosa, 76, 20018 - San Sebastián, Spain
[4]School of Information and Communication Technology, KTH Royal Institute of Technology, E229, 16440 Kista, Sweden

Corresponding author: max.lemme@uni-siegen.de

**Abstract**

Graphene has extraordinary mechanical and electronic properties, making it a promising material for membrane based nanoelectromechanical systems (NEMS). Here, chemical-vapor-deposited graphene is transferred onto target substrates to suspend it over cavities and trenches for pressure-sensor applications. The development of such devices requires suitable metrology methods, i.e., large-scale characterization techniques, to confirm and analyze successful graphene transfer with intact suspended graphene membranes. We propose fast and noninvasive Raman spectroscopy mapping to distinguish between freestanding and substrate-supported graphene, utilizing the different strain and doping levels. The technique is expanded to combine two-dimensional area scans with cross-sectional Raman spectroscopy, resulting in three-dimensional Raman tomography of membrane-based graphene NEMS. The potential of Raman tomography for in-line monitoring is further demonstrated with a methodology for automated data analysis to spatially resolve the material composition in micrometer-scale integrated devices, including free-standing and substrate-supported graphene. Raman tomography may be applied to devices composed of other two-dimensional materials as well as silicon micro- and nanoelectromechanical systems.

Since its discovery, graphene has received substantial attention due to its exceptional electrical and mechanical properties. It is the thinnest and one of the strongest materials known with a thickness of 0.34 nm (one atomic layer) and a Young's modulus possibly exceeding 1 TPa.[1] These properties make it possible to suspend this intrinsically nanoscale material across several hundreds of micrometers distance.[2] Furthermore, it is stretchable up to 20% of its length[3] and impermeable to gases, including Helium.[4] In addition, it has a high carrier mobility[5,6] and exhibits a piezoresistive effect i.e. the resistance changes when mechanical stress is applied, exhibiting a gauge factor of the order of 3.[7] These properties suggest a great variety of possible applications, in particular in the field of micro- and nanoelectromechanical systems (MEMS/NEMS). Demonstrations reported in literature include graphene cantilever beams for nano-switches,[8] mechanical graphene radio-frequency (RF) switches,[9] desalination using nanoporous graphene membranes[10] or highly scalable graphene pressure sensors.[7,11] All of these applications require suspended (single or multiple layer) graphene over either closed cavities or open trenches. Mechanically exfoliated graphene has been used extensively to investigate fundamental material properties and to demonstrate prototype devices.[12–16] However, mechanically exfoliated flakes, reaching typical average sizes of only a few square micrometers, are further distributed randomly on the substrate with a varying number of layers. Even though precise placement of such graphene flakes over cavities and holes is feasible, it requires continuous monitoring of the transfer process under an optical microscope. This makes the process very cumbersome, and only prototype devices can be demonstrated. Large-scale fabrication of NEMS devices with high yield obviously requires suspended graphene regions over large-areas. Chemical-vapor deposition (CVD) of graphene on catalytic metallic substrates and subsequent transfer, however, is a promising technique,[17] even for suspended graphene films.[18] Here, the graphene size is only limited by the growth substrate.[19]

Copper (Cu) is typically used as a substrate to grow graphene as it is compatible with silicon technology and relatively inexpensive, in particular if it is reused.[20] The graphene can then be transferred on large-scale, opening up the potential wafer-scale processing of graphene-based devices.[19,21] Regardless of the growth and transfer methods, the presence of graphene on the device substrates is most commonly confirmed by Raman spectroscopy, which is a fast and potentially non-invasive analytical method for graphene and other twodimensional (2D) materials.[22] Raman spectroscopy can provide a wide range of information about graphene such as the number of layers[22], stacking order[23], quality[22], defects[24,25], doping[24,26,27], thermal properties[28,29] and strain.[30,31] These can be measured and analyzed from substrate supported graphene (SSG)[22] as well as suspended, free-standing graphene (FSG)[32], which is otherwise not easy to detect due to its thinness and its high broad-band transparency.[33] Most Raman measurements are obtained by recording single spectra at a few points of interest in the graphene. For large-scale characterization, this acquisition process can be cumbersome and may yield random results. Therefore, large-area scans are preferred over single spectra opening up the possibility to analyze a continuous area where each pixel of the resulting image corresponds to one single spectrum. Gajewski *et al.*[34] and Suk *et al.*[17] demonstrated high resolution Raman spectroscopy to investigate strain on FSG versus SSG using small area maps. With wide-field Raman spectroscopy it is possible to map large areas with short acquisition times.[35] These systems, however, need one or multiple filters for predefining the desired Raman peaks and are therefore not able to record complete spectra. Furthermore, the x, y and z-resolution is in the micrometer range[35] in comparison to the sub-micrometer resolution of the conventional Raman systems. Three-dimensional (3D) Raman tomography is a very useful feature of Raman spectroscopy where several line or area scans are recorded at different depths (z-positions) through a material. This technique has been used for observing composition of biological samples[36–38], lattice defects[39] and phase transitions[40] in semiconductors and quality of synthetic diamonds[41] in the past.

    Here, we demonstrate a fast, noninvasive, and large-area confocal Raman spectroscopy and tomography method that distinguishes SSG and FSG rapidly and on a large-scale. The method can potentially be integrated into in-line processing for rapid, large-scale metrology. It utilizes single spectra, 2D area mapping, cross-sectional scans as well as 3D mapping of graphene. By combining the areal mapping data with the depth information, a true 3D image of a specific volume can be reconstructed, and the in-plane continuity and homogeneity of graphene with a resolution of 300 nm in lateral and 900 nm in vertical dimensions can be evaluated. Furthermore, we present 3D Raman tomographic scans of FSG areas and their

surroundings. With this technique, one can investigate the surface and the inside of a material, i.e., material layers, different phases, possible inclusions, and cavities, which otherwise cannot be retrieved in a nondestructive way with conventional analytical techniques. In this work, we apply this method as an efficient way to screen whether graphene membranes are damaged or destroyed during device fabrication. In principle, the method can be extended to investigate other electron devices, particularly those including Raman active 2D materials.

Test structures for FSG membranes were fabricated on Si substrates with 1.6 µm of thermally grown silicon dioxide (SiO2, Figure 1a). Cavities were then etched 1.4 µm deep into the SiO2 with a reactive ion etching (RIE) process (Figure 1b). The test areas contained tightly spaced cavities including round holes with sizes between 3 and 15 µm diameter as well as square and rectangular geometries with similar sizes. Finally, CVD graphene grown on copper foils was transferred with an optimized transfer process (Figure 1c). Figure 1d,e shows micrographs of the test structure areas on the chip after graphene transfer.

Raman experiments were conducted using a WITec apyron confocal Raman microscope equipped with a 532 nm laser coupled to the microscope using a single-mode optical fiber. For the high reproducibility of the experiments, the laser power was determined using a power meter integrated in the optical path and controlled by the TruePower functionality of the apyron microscope. The microscope was connected to a 600 mm ultrahigh-throughput spectrometer (UHTS 600) using a photonic fiber. A 300 g/mm grating was used for the dispersion and an EMCCD camera for the detection of the scattered light. For best vertical and lateral resolution as well as for best collection efficiency, a 100× objective with a numerical aperture (NA) of 0.9 was chosen for the measurements. It should be noted that we have used a low-resolution 300 g/mm grating in combination with the 600 mm spectrometer to obtain moreintense signals with a pixel resolution of 2.2 cm−1 @ 2800 relative cm−1. This, in turn, allowed us to decrease the scan times significantly and to quickly identify the FSG regions over large-areas. However, after quick identification of suspended graphene membranes using our methodology, one can use a high-resolution grating like 1800 g/mm or higher to obtain more precise quantitative information about desired Raman peak positions and peak widths at those specific locations. The samples were positioned either using a motorized stepper stage for the large-area images or using a piezoelectric stage for the high-resolution images. The z position was controlled by a stepper motor stage with 10 nm single step accuracy. A schematic image of the Raman setup is shown in Figure 2a.

Typical Raman spectra of graphene show three characteristic features, namely the D peak at ~ 1350 cm$^{-1}$, the G peak at ~ 1580 cm$^{-1}$ and the 2D peak at ~ 2700 cm$^{-1}$.[22] The intensity, position, and full width at half maximum (fwhm) of these peaks depend on the quality, the number of layers, the defect density of graphene,[22] doping[24,26,27] and strain.[29,30] Also, the incident laser power can have significant effects on graphene and cause irreversible damage to the graphene structure.[42,43] Since the threshold for each material, each device configuration, and each Raman system setup is different, the effect of the laser power has been determined before the actual measurements. The limit of the non-invasive regime was established by analyzing the 2D peak of SSG and FSG, as this is the most intense peak for single layer graphene and can be easily observed even with very low laser powers. Furthermore, it is a very good indication for the presence of graphene.[22] The laser power was varied over a wide range from 0.1 mW to 10 mW and single Raman spectra of SSG and FSG on the Si/SiO$_2$ substrates were recorded. The peak position and FWHM of the 2D peak are plotted as a function of the laser power in in Figure 2b.

A significant red shift (shift toward lower energies) of the 2D peaks was observed as well as an increase in the fwhm for both SSG and FSG samples for laser powers above 1 mW. At low or moderate laser powers, a reversible photoinduced doping of SSG takes place, which results in a reversible 2D peak shift.[44] However, for higher laser powers, there is a significant rise in the sample temperature, which can result in an annealing effect causing a nonreversible shift and broadening of the 2D peak.[42] It is in this regime that Raman microscopy becomes invasive, as it modifies the properties of the graphene during the measurements. Therefore, a laser power of 1 mW was used for all the Raman measurements in the present study to remain in the noninvasive regime. The difference in the 2D peak positions between SSG and FSG can be due to the substrate influence, i.e, strain[45] and doping conditions.[34,46,47]

The optical images in Figure 3a,b provide an overview of the entire chip and the measured test area. Figure 3b shows a high density of cavities with different sizes and geometries in which the goal was to identify intact membranes. Such a fast scan is useful to determine where a more-elaborate and time consuming Raman tomography scan makes sense or is necessary or if interesting results can be expected. To compensate for slight tilts of the sample relative to the scanning stage, a tilted plane was defined using the system's software and the sample was scanned along this plane, keeping the sample in focus at all times. A fast large-area Raman scan of 530 × 530 µm$^2$ was conducted at 20 ms per spectrum on the test structure area resulting in a total scan time of 13 min. 200 × 200 spectra were taken on this area, and an image was constructed in which each pixel is represented by one single spectrum. The resulting spectra were filtered for cosmic rays and baseline-corrected. Finally, the intensity of the 2D peak was extracted from each acquired spectrum and plotted as a map in Figure 3c. The areas with FSG, indicated by a high intensity of the 2D peak, can easily be identified from this Raman image and appear as bright yellow areas. The analysis of the 2D peak region thus provides a clear distinction between covered and noncovered cavities. This was confirmed with a high-resolution Raman area scan of a 200 × 150 µm$^2$ area (Figure 3d).

The scanning Raman technique provides additional information that can be extracted after data acquisition. We scanned a 35 × 32 µm$^2$ area with four cavities in 0.5 µm wide steps and 100 ms integration time per pixel. The cavities' location on the chip is indicated with a white square in Figure 3c. The resulting intensity map is plotted in Figure 4a after the removal of the cosmic ray and background signals. The spectra shown in Figure 4b were derived by cluster analysis[48] and spectral de-mixing[48], a software algorithm that automatically detects similar regions in the spectral data that allows to objectively identify different materials (clusters) in a Raman spectrum (Figure 4b). The k-means cluster analysis is described in detail in section S2 of the Supporting Information. Using these spectra, a basis analysis was performed on the full data set, revealing the areas in which the different spectra can be found. The spectra and the images in Figure 4 are color-coded and thus serve to determine and assign the presence of FSG (yellow), SSG (turquoise), amorphous carbon (magenta), and silicon (blue). Furthermore, a very low D peak can be seen indicating a rather high quality of the graphene with only few defects. A black background color displays the absence of any of the filtered materials (i.e., air or silicon dioxide). From the resulting image, one can determine that two of the cavities (cavities no. 3 and 4) are completely covered, one is partly covered (no. 1), and the fourth (no. 2) is not covered at all.

In addition to this area scan and material classification, we performed depth scans in the z direction along a 35 µm line across cavity 2 and 3, indicated by the green line in Figure 4a. These scans started with the focal point above the sample surface and, after finishing each line scan, the focal point of the laser was automatically adjusted in the z direction so that it scanned the sample step-by-step into the material. The maximum z resolution of the Raman system is 900 nm, and a total of 24 line scans with a step of 300 nm in the z direction were recorded with an integration time of 500 ms per pixel. The maximum theoretically possible z resolution for a green laser was calculated to be between 630 and 680 nm.[49] In addition, oversampling can be used to improve the image quality, which was done in case of the cross-sectional scan. For materials with different Raman fingerprints, a detectable chemical contrast peak-by-peak analysis is possible. Further improvements in resolution down to a maximum of 200 nm in lateral and 600 nm in the z direction can be accomplished by using high-NA objectives, e.g., oil immersion objectives in combination with a different laser. The cross-sectional image (Figure 4c) was reconstructed from these measurements and shows the SSG turquoise on top of the Si (blue). The FSG in yellow on cavity 3 is also clearly visible in comparison to cavity 2, which is not covered with a graphene membrane. In particular, the thickness of graphene is not displayed to scale due to the limited z resolution. At the top of the cavity, an area of amorphous carbon (magenta) at the periphery of the cavities can be seen, indicated by a broad peak with overlapping D and G peaks with no signal of the 2D peak. Such overlapping or broad peaks around 1300–1700 cm−1 are characteristic of amorphous or polymeric carbon products.[50] In the present case, there is a strong likelihood of polymeric residue formation from the RIE cavity etch process with $CHF_4$ gas, which is known to create a $C_xF_y$ layer on the substrate surface,[51] hence the observed Raman feature. A rise in the D peak can be observed in the FSG (yellow), SSG (turquoise), and the amorphous carbon (magenta) spectra. It can be attributed to the formation of amorphous carbon during the 450 °C annealing process that was used to remove

PMMA after transfer. The D peak in the FSG spectrum seems to have lower intensity in comparison to the spectra obtained in other regions. This is due to the fact that the presented spectra are not normalized, and the 2D peak for FSG is much more intense, as discussed earlier, thus suppressing the D peak intensity. The 1.6 µm thick SiO2 layer is displayed as a black gap between the graphene and the Si. Note that the Si layer at the bottom of the cavities appears to be shifted in the z direction compared to that below the SiO2 in the crosssectional images (Figure 4c), even though the layer is flat in reality and the cavities do not reach into the Si layer. This is an artifact due to different optical paths of the laser in both cases: the residual $SiO_2$ thickness of approximately 200 nm in the cavities is much thinner than that on the rest of the substrate (1.6 µm). The $SiO_2$ is optically transparent for the wavelength of 532 nm. During the measurement on the Si, the beam is focused at a point located inside the Si layer. Therefore, the incident and the outgoing light passing through the $SiO_2$ layer are refracted at the interfaces between air and $SiO_2$ and at the interface between $SiO_2$ and Si. As a result, the focal point appears closer to the detector than it actually is. This effect becomes stronger for increasing $SiO_2$ thickness, which results in the Si layer being displayed at different z-positions. For known beam paths, device designs, or structures, such artifacts can be removed by including an appropriate algorithm into the analyzing software.

A 3D tomography image of the graphene membrane test structure was reconstructed from eight area scans at different z positions with 300 nm step width (Figure 4d). The total scan time is approximately 2.5 h, as eight images of 18 × 18 µm$^2$ with 3600 single spectra and 300 ms integration time each were taken to achieve a high-resolution image. Different materials were again color-coded according to Figure 4b. The resulting 3D image (Figure 4e) allows us to analyze the entire device including materials inside the substrate. For example, Figure 4e clearly shows that the graphene is indeed suspended across the cavity and does not extend into the cavity (within the limited z resolution). Furthermore, no graphene is present at the bottom of the holes. The ring of amorphous carbon (magenta) is mainly detected at the top edge of the cavity and is very slightly visible in the area where the sidewalls meet the bottom of the cavity but not along the sidewalls and the bottom.

One or several layers of the 3D-image stack can be analyzed in depth by extracting detailed material information at different locations in the 3D space. In case of the graphene, the doping and strain conditions in FSG and SSG are of great interest and can be determined by choosing the second image from the stack show in Figure 4d. Figure 5a shows the representative Raman spectra for suspended and nonsuspended regions, with positions indicated in Figure 5b corresponding to the gray, purple, and pink crosses. A lower D peak and an increase in G and 2D peak intensity from SSG (gray and pink spectra) to FSG (purple spectra) can be observed. Area maps in Figure 5b–e provide additional detailed information regarding the distribution of the displayed values, which cannot be obtained with single spectra recorded on just a few places across the sample. Figure 5b shows a clear increase in 2D peak intensity from SSG to FSG. In addition, the map of the peak position (Figure 5c) and fwhm (Figure 5d) of the 2D peak have been reconstructed using a Lorentzian fit of that peak. Black pixels in the images correspond to zero values and indicate spectra in which no solution for the automated fitting could be found. Finally, the map of the 2D over G peak intensity ratio is displayed in Figure 5e. The low D peak, the high and narrow 2D peak, and a higher 2D intensity as well as a higher intensity ratio of 2D and G peaks in the case of FSG indicate different graphene properties than SSG.[22,23,52] Further analysis of the available data shows that the 2D peaks are blue-shifted (shifts toward higher energies) by approximately 15 cm−1 in the SSG. This shift can also be observed in the laser-power-dependent Raman measurements (Figure 2). The observed decrease in I2D/IG, due to a decrease in 2D peak intensity in SSG (Figure 5e), can be attributed to hole doping.[34,46,47] The blue shift of the 2D peak in SSG results from the $SiO_2$ substrates, which is known to result in p-doping due to trapped charges in the oxide[53] and ambient molecular doping / humidity[54]. FSG over cavities is likely to have built-in tensile strain due to the geometry of membranes, i.e. slight sagging of the membranes or strain which was introduced during processing.[45] In contrast, thermal annealing is known to induce compressive strain in SSG on $SiO_2$ due to thermally induced mechanical transformations.[55] Even though both SSG and FSG show spectral inhomogeneity across the regions, FSG exhibits a clear contrast to the surrounding supported regions. Therefore, different doping and strain levels are likely responsible for the observed distinct Raman characteristics of FSG and SSG.

A similar device to the one discussed in Figures 4 and 5 was investigated with Raman tomography. Here, the cavities were covered with polytetrafloerethylen (PTFE), and the results are shown in Figure S1. Here, the PTFE layers on the sidewalls clearly show up at the sidewalls down to the bottom of the cavity. This image demonstrates that also underneath the graphene membrane, i.e., chemical information can be retrieved noninvasively from the inside of devices. This extends the applicability of the Raman tomography method to graphenecovered chemical micro reactors[56] and micro- and nanobubbles covered by graphene[57,58] or microfluidic systems. Si-based devices could be investigated with lasers with higher wavelengths such as 785 or 1064 nm for higher penetration, albeit with the disadvantage of decreased resolution. Here, applications in silicon MEMS can be envisioned, to determine noninvasively the full release of MEMS systems or the status of integrated and covered parts like membranes or the supporting structures. Another application example could be vertical heterostructures like graphene-based hot electron transistors,[59] even though in all cases, the limited z-resolution prevents the determination of exact thicknesses and locations of very thin layers in the devices. To determine these, destructive methods such as cross-sectional SEM or TEM still have to be used.

In summary, we have demonstrated a noninvasive, large-area, and relatively rapid metrology method to determine the presence of free-standing CVD graphene and to distinguish it clearly from graphene supported on a substrate based on confocal Raman microscopy. The noninvasive limit of the laser power was determined in a wide power range. Specific regions of a test chip were analyzed by single-spectrum, high-resolution 2D area mapping. We further introduced a novel methodology to reconstruct 3D tomographic images of solid-state devices, which are a unique way to gain nondestructive insight with fairly high spatial resolution of the structure, material, compound, and other phenomena. The stepper motor and piezo scanner in the Raman system allow the correlation of large-area-scale data with high-resolution information. The present study suggests, in principle, in-line integration of a noninvasive and noncontact metrology method for graphene as well as other devices composed of Raman-active materials. This would be of high interest for future applications as it provides a means of quality control during fabrication, which could be automated to investigate hundreds of points of interest across a chip. The method is suitable to provide chemical and structural information not only of the surface but also inside devices and thus enables the noninvasive analysis of a wide range of applications, including silicon MEMS and microfluidics.

**Methods**

The test structures were fabricated on a p-doped silicon substrate with a 1.6 µm thermal oxide layer. The structures are cavities with diameters between 3 and 20 µm with circular, square, and rectangular geometry. They were etched 1.4 µm deep into the SiO2 layer using a photoresist mask and an Ar- and CHF3-based reactive-ion-etching (RIE) process using 200 mW and 40 mTorr for a vertical etch profile. The monolayer graphene samples were grown by chemicalvapor deposition (CVD) in a 4 in. cold walled reactor (Aixtron BM) using copper (Cu) foil as the growth substrate. Cu surface pretreatment and high-temperature annealing were performed before the graphene growth. The graphene synthesis was carried out using methane as the carbon source. After the synthesis, a semidry proprietary transfer process using PMMA as the supporting layer was used to suspend the monolayer graphene onto the cavities. The supporting polymer was removed by annealing the sample at 450 °C for 2h in N2 atmosphere. Finally, the graphene samples were annealed at a temperature of 450 °C to minimize residues formed by processing.

**Acknowledgements**

Funding from the German Federal Ministry of Education and Research (BMBF, NanoGraM, 03XP0006), the European Research Council (ERC, InteGraDe, 3017311), the German Research Foundation (DFG LE 2440/1-2) and he Spanish Ministry of Economy and Competitiveness (MINECO, NanoGraM, CDTI IDI-20150827 and SPRI IG-2015/0001026) is gratefully acknowledged.

**Supporting information**

- Raman 3D-image stack and tomography image of a second device fabricated using a PTFE layer to coat the sidewalls of the cavity
- Description of the k-means cluster analysis method used in the software to analyze the Raman data

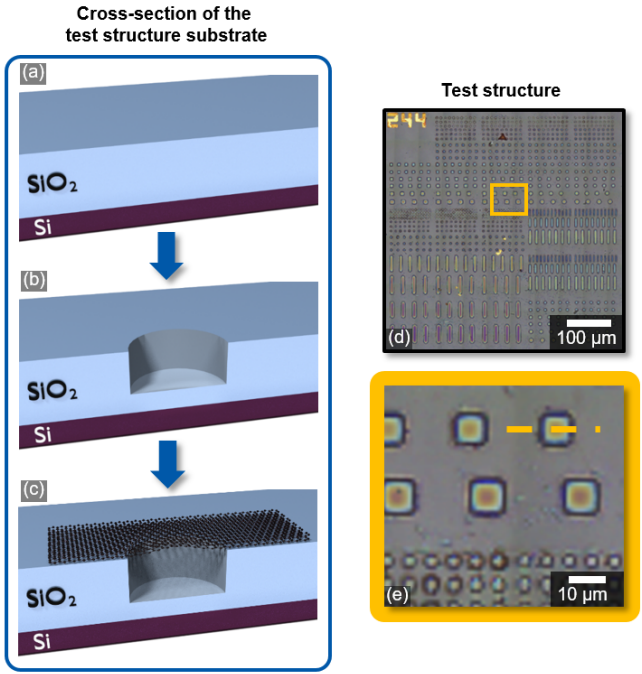

*Figure 1: Cross-sectional schematic of the fabrication process for test structures: (a) substrate; (b) RIE cavity etch; (c) graphene transfer onto test structures to cover the cavities. (d) Optical micrograph of the test structures; (e) zoomed-in area of the red square marked in (d) indicating the cross-sectional view of the fabrication process.*

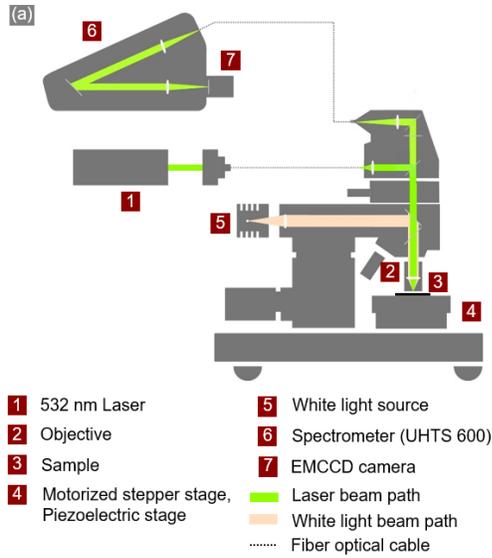 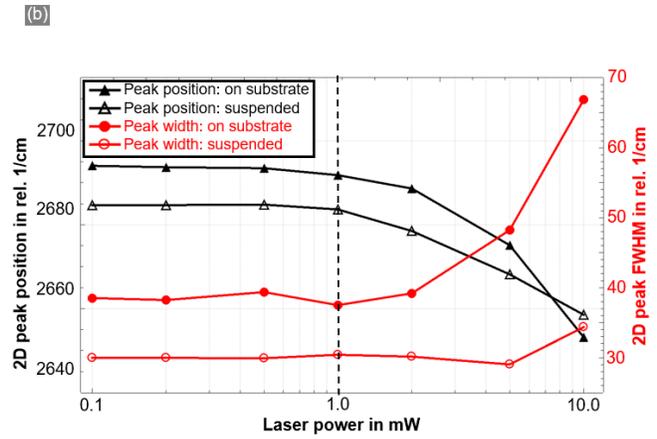

*Figure 2: (a) Typical beam path of an α300/apyron system; (b) Raman peak position and the full width half-maximum (fwhm) of the 2D peak as a function of laser power for FSG and SSG. Values for FSG are higher than for SSG due to substrate influence. The dashed vertical line indicates the limit for the noninvasive regime for characterizing graphene.*

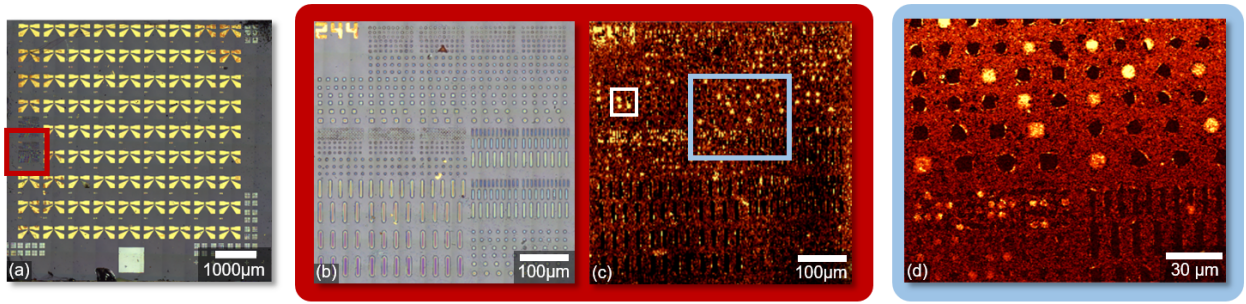

*Figure 3: (a) Stitched white-light image of the complete chip; (b) stitched white-light image of the test structures; (c) Raman image of the test structures displaying 2D peak intensity; (d) zoomed-in image of the large-area map (light blue square) showing the 2D peak intensity for an area with covered and uncovered cavities.*

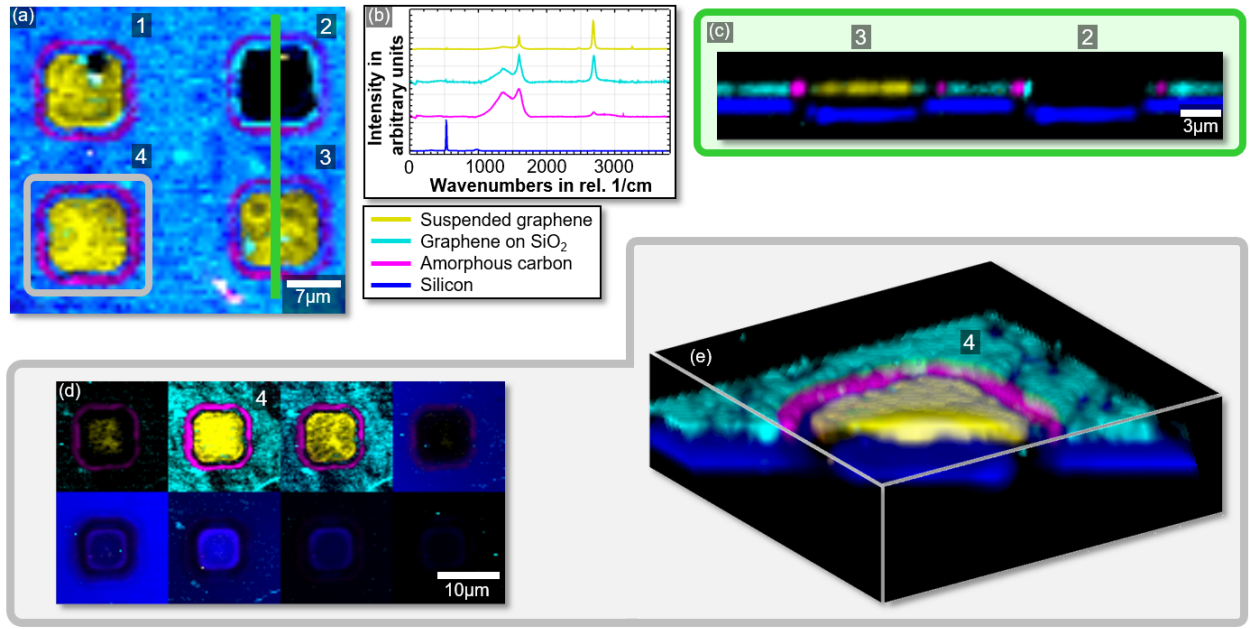

*Figure 4: (a) 2D Raman scan of four cavities (1, 2, 3, and 4) from the test structures indicated by a white square in Figure 3c; (b) demixed spectra showing FSG (yellow), SSG on SiO2 (turquoise), amorphous carbon (magenta), and silicon (dark blue); (c) cross-section of cavities 2 and 3 taken at the green line indicated in panel (a) with a gray square; (d) sequence images of cavity 4 for the 3D scan taken at different z-positions; (e) 3D tomography image of cavity 4 reconstructed from the data (gray square in panel (a)).*

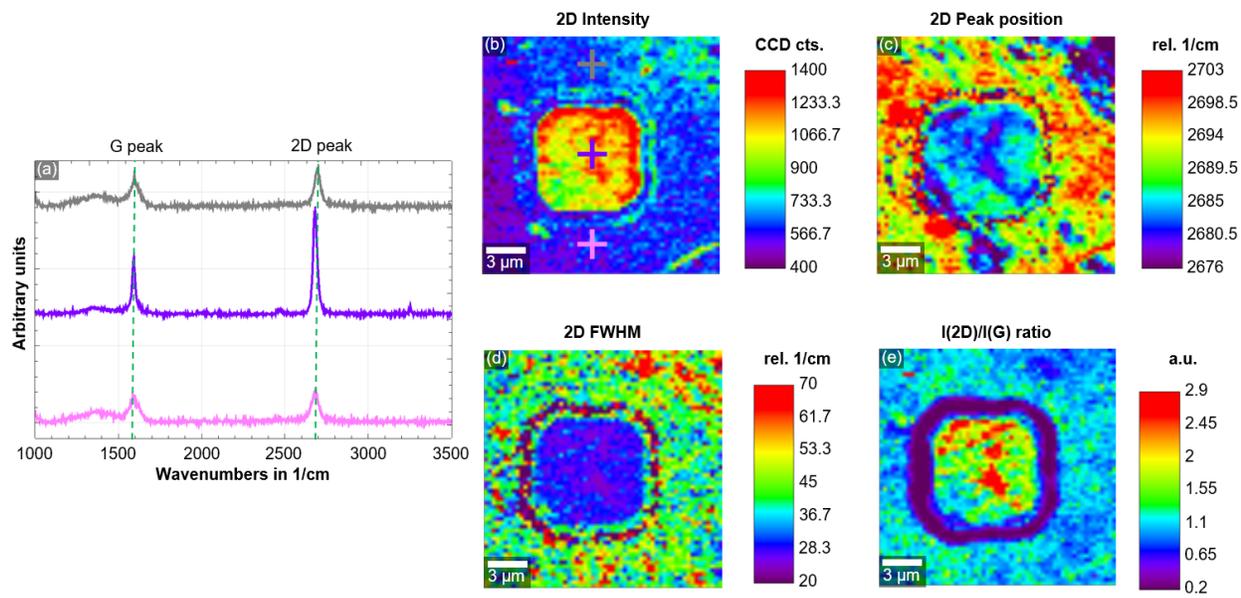

Figure 5: (a) Single spectra of the surrounding SSG (black and pink) and the FSG in the middle (purple); (b) intensity map of the 2D peak with crosses marking the single spectra in panel (a); (c) 2D peak position; (d) the fwhm of the 2D peak and (e) $I_{2D}/I_G$ ratio.

Graphic for the Table of Contents (TOC)

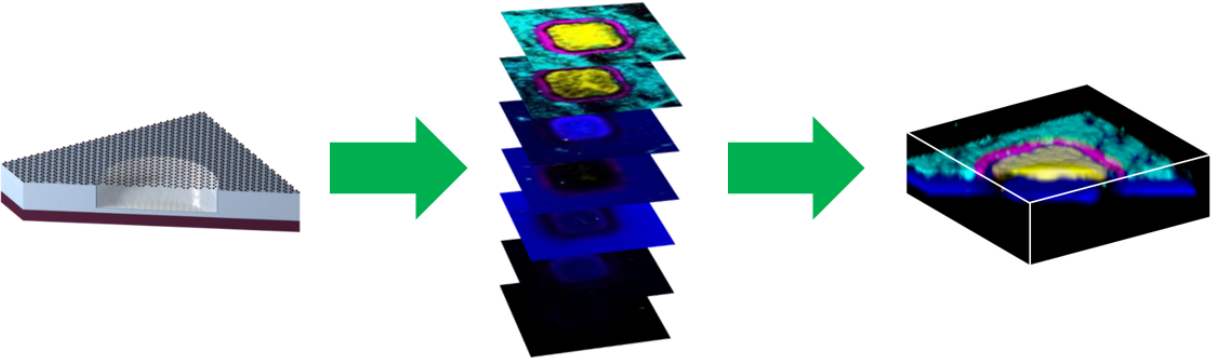